\documentclass[applemac]{aa}
\usepackage{natbib}
\usepackage{txfonts}
\usepackage{graphicx}
\usepackage{color}
\bibpunct{(}{)}{;}{a}{}{,}

\begin{document}

\title{Modeling deuterium chemistry in starless cores: full scrambling versus proton hop}
\author{O. Sipil\"a\inst{1},
		P. Caselli\inst{1},
                and J. Harju\inst{2,1}
}
\institute{Max-Planck-Institute for Extraterrestrial Physics (MPE), Giessenbachstr. 1, 85748 Garching, Germany \\
e-mail: \texttt{osipila@mpe.mpg.de}
\and{Department of Physics, P.O. Box 64, 00014 University of Helsinki, Finland}
}

\date{Received / Accepted}

\abstract
{We constructed two new models for deuterium and spin-state chemistry for the purpose of modeling the low-temperature environment prevailing in starless and pre-stellar cores. The fundamental difference between the two models is in the treatment of ion-molecule proton-donation reactions of the form $\rm XH^+ + Y \longrightarrow X + YH^+$, which are allowed to proceed either via full scrambling or via direct proton hop, i.e., disregarding proton exchange. The choice of the reaction mechanism affects both deuterium and spin-state chemistry, and in this work our main interest is on the effect on deuterated ammonia. We applied the new models to the starless core H-MM1, where several deuterated forms of ammonia have been observed. Our investigation slightly favors the proton hop mechanism over full scrambling because the ammonia D/H ratios are better fit by the former model, although neither model can reproduce the observed $\rm NH_2D$ ortho-to-para ratio of 3 (the models predict a value of $\sim$2). Extending the proton hop scenario to hydrogen atom abstraction reactions yields a good agreement for the spin-state abundance ratios, but greatly overestimates the deuterium fractions of ammonia. However, one can find a reasonably good agreement with the observations with this model by increasing the cosmic-ray ionization rate over the commonly-adopted value of $\sim$$10^{-17}\,\rm s^{-1}$. We also find that the deuterium fractions of several other species, such as $\rm H_2CO$, $\rm H_2O$, and $\rm CH_3$, are sensitive to the adopted proton-donation reaction mechanism. Whether the full scrambling or proton hop mechanism dominates may be dependent on the reacting system, and new laboratory and theoretical studies for various reacting systems are needed to constrain chemical models.}


\maketitle

\section{Introduction}\label{s:introduction}

Ion-molecule proton-donation reactions are at the heart of low-temperature chemistry in the interstellar medium (ISM) \citep{Herbst73}. One particular molecule that is formed through such reactions is ammonia, which is an important diagnostic tool for probing the gas temperature in the ISM, and has been extensively used for this purpose \citep[e.g.,][and references therein]{Friesen17}. Ammonia is a relatively simple molecule and its main gas-phase formation and destruction pathways have long since been identified: its formation begins with the slightly endothermic hydrogen atom abstraction reaction
\begin{equation}\label{eq:ammonia1}
\rm N^+ + H_2 \longrightarrow NH^+ + H \, ,
\end{equation}
followed by the analogous reactions
\begin{align}
\rm NH^+ + H_2 &\longrightarrow \rm NH_2^+ + H \label{eq:ammonia2} \\
\rm NH_2^+ + H_2 &\longrightarrow \rm NH_3^+ + H \label{eq:ammonia3} \\
\rm NH_3^+ + H_2 &\longrightarrow \rm NH_4^+ + H \label{eq:ammonia4} \, .
\end{align}
The dissociative recombination of $\rm NH_4^+$ with free electrons then forms ammonia. The most important destruction pathway ($\rm H_3^+ + NH_3 \longrightarrow NH_4^+ + H_2$) actually leads back to the ammonia formation ladder.

Molecules with multiple hydrogen (or deuterium) atoms can exist in two or several forms owing to the nuclear spin of the proton  (deuteron). For example for $\rm H_2$, the combination of two protons gives rise to two spin states where the nuclear spin wavefunction is either antisymmetric (para-$\rm H_2$; rotational quantum number $J = 0,2,4...$) or symmetric (ortho-$\rm H_2$; $J = 1,3,5...$); these states behave as distinct chemical species that (practically) cannot be converted from one form to the other by radiation or by inelastic collisions. The $\sim$170\,K energy difference between the $J = 0$~and~1 rotational levels is an important source of chemical energy in dense clouds, and allows in particular the exothermic reaction
\begin{equation}
\rm H_3^+ + HD \longleftrightarrow H_2D^+ + H_2 + 232\,K
\end{equation}
to proceed also in the backward reaction near 10\,K. $\rm H_2D^+$ is the main driver of deuteration in cold clouds \citep[e.g.][]{Rodgers01}. Spin-state and deuterium chemistry are thus very closely tied, and to understand observations of deuterated species through chemical modeling, one needs to take spin-state chemistry into consideration. Several such chemical models have been introduced in the past decades, describing gas-phase or gas-grain chemistry \citep[e.g.,][]{Pagani92,WFP04,Sipila10,Furuya15,Roueff15,Hily-Blant18}.

$\rm NH_3$ can exist in a total of nine variants when one takes deuterium and spin states into account. Almost all of these exhibit emission lines that are observable from the ground. Deuterated ammonia traces the innermost areas of starless cores, where many otherwise common molecules such as CO are frozen onto dust grains, as demonstrated in the case of H-MM1 by \citeauthor{Harju17a}\,(\citeyear{Harju17a}; hereafter H17) who carried out an extensive modeling exercise in order to reproduce the observed D/H and spin-state abundance ratios. Their model however failed to reproduce in particular the ortho-to-para (hereafter o/p) ratios of $\rm NH_2D$ and $\rm NHD_2$, and it remains unclear why the model prediction deviates from the observed values.

H17 used in their study the gas-grain chemical model of \citet{Sipila15b} that includes a self-consistent description of spin-state chemistry for multiply-deuterated species, derived using group theory. The underlying assumption in that model is that the relevant chemical reactions proceed through full scrambling (FS), where it is believed that the reactants form a relatively long-lived intermediate complex in which several atom interchanges can take place \citep{Gerlich92}. FS actually consists of three possible outcomes \citep{Oka04}: identity, proton hop (PH), and proton exchange. It has been shown that the reactions belonging to the $\rm H_3^+ + H_2$ system proceed preferentially through FS even at low temperature \citep{Hugo09,Suleimanov18}. It is however not at all clear that other reacting systems involving more massive molecules also prefer FS over PH at low temperature; \citet{LeGal17} have shown that the reactions relevant to $\rm H_2Cl^+$ formation in diffuse cloud conditions do in fact prefer PH, but overall there is unfortunately very little data on this topic in the literature.

In this paper we investigate the effect of proton exchange on ion-molecule chemistry in the ISM, by modifying the chemical model of \citet{Sipila15b} by allowing proton-donation reactions, in which proton exchanges were previously allowed, to proceed instead through PH only. From here on, we simply call these two types of model either the FS or PH model, respectively. We concentrate our analysis on deuterated ammonia and reproduce some of the modeling results of H17, using our gas-grain chemical model in combination with radiative transfer simulations of ammonia emission lines. We also investigate the effect of the reaction mechanism on other species besides ammonia, with the aim of quantifying whether FS or PH is preferred. A small part of the results presented here have already been introduced in \citet{Caselli19}.

The paper is organized as follows. In Sect.\,\ref{s:model}, we introduce the new FS and PH chemical networks, and discuss the details of our modeling. Section~\ref{s:results} presents the results of our analysis, which are further discussed in Sect.\,\ref{s:discussion}. We give our conclusions in Sect.\,\ref{s:conclusions}.

\section{Model}\label{s:model}

\subsection{FS versus PH: deuteration}\label{ss:scramblingVersusHopDeuteration}

We consider in the FS model that reactions proceed through an intermediate complex that constitutes a pool of atoms from which the atoms are drawn one by one. The branching ratios are constructed by multiplying simple probabilities. For example in the reaction $\rm NH_3 + D_2H^+ \longrightarrow (NH_4D_2^+)^* \longrightarrow NH_3D^+ + HD$, the intermediate complex contains four H atoms and two D atoms. The four H/D atoms required to form $\rm NH_3D^+$ can be picked in ${4 \choose 1} = 4$ equivalent combinations: HHHD, HHDH, HDHH, and DHHH. The individual probabilities for each combination are the same, for example for HHHD: $\frac{4}{6} \frac{3}{5} \frac{2}{4} \frac{2}{3} = \frac{2}{15}$. So, in total, the branching ratio for $\rm NH_3D^+$ formation is ${4 \choose 1} \frac{2}{15} = \frac{8}{15}$. There is only one combination that forms $\rm NH_4^+$ (probability $\frac{1}{15}$), and so the probability to form $\rm NH_2D_2^+$ is $\frac{6}{15}$.

The PH mechanism instead assumes that atom exchanges do not occur during the reaction, and the donating ion simply transfers one proton or deuteron to the other reactant; atom-exchanging reaction channels are closed off. Therefore the probability to form $\rm NH_3D^+$ in $\rm NH_3 + D_2H^+$ is simply $\frac{2}{3}$, for example. We obtain in general fewer product pathways when assuming PH instead of FS.

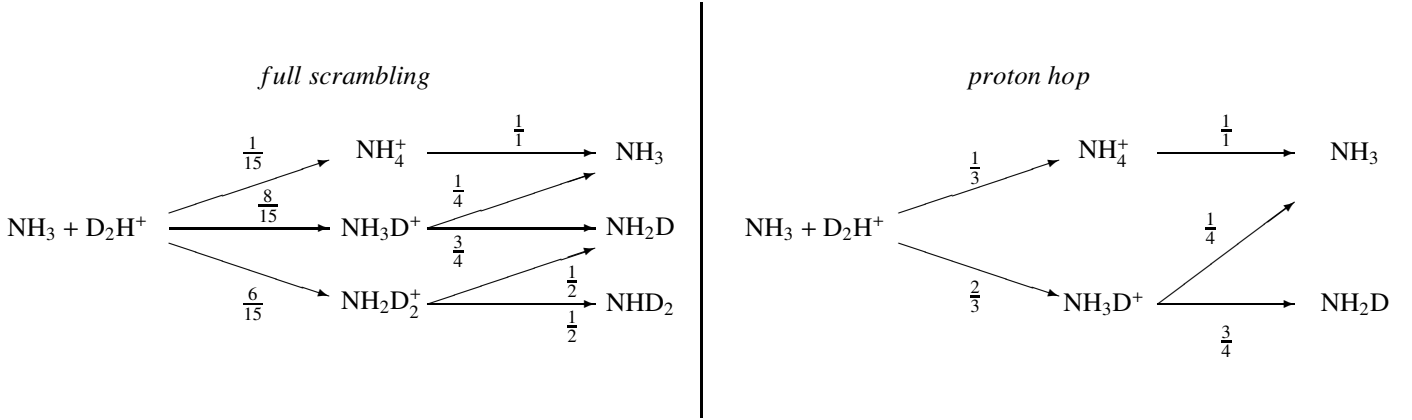
\begin{figure*}[hbt]
\unitlength=1.0mm
\begin{picture}(160,70)(0,10)

\color{black} 


\put(45,60){\makebox(0,0){$\sl full~scrambling$}}

\put(10,40){\makebox(0,0){$\rm NH_3 + D_2H^+$}}

\put(50,50){\makebox(0,0){$\rm NH_4^+$}}
\put(50,40){\makebox(0,0){$\rm NH_3D^+$}}
\put(50,30){\makebox(0,0){$\rm NH_2D_2^+$}}

\put(84,50){\makebox(0,0){$\rm NH_3$}}
\put(84,40){\makebox(0,0){$\rm NH_2D$}}
\put(84,30){\makebox(0,0){$\rm NHD_2$}}

\put(22,42){\vector(3,1){21}}
\put(22,40){\vector(1,0){21}}
\put(22,38){\vector(3,-1){21}}
\put(33,50){\makebox(0,0){$\frac{1}{15}$}}
\put(35,43){\makebox(0,0){$\frac{8}{15}$}}
\put(33,30){\makebox(0,0){$\frac{6}{15}$}}

\put(56,50){\vector(1,0){22}}
\put(56,40){\vector(1,0){22}}
\put(56,40){\vector(3,1){22}}
\put(56,30){\vector(1,0){22}}
\put(56,30){\vector(3,1){22}}
\put(68,53){\makebox(0,0){$\frac{1}{1}$}}
\put(75,27){\makebox(0,0){$\frac{1}{2}$}}
\put(75,33){\makebox(0,0){$\frac{1}{2}$}}
\put(60,37){\makebox(0,0){$\frac{3}{4}$}}
\put(60,45){\makebox(0,0){$\frac{1}{4}$}}


\put(135,60){\makebox(0,0){{$\sl proton~hop$}}}

\put(107,40){\makebox(0,0){$\rm NH_3 + D_2H^+$}}

\put(145,50){\makebox(0,0){$\rm NH_4^+$}}
\put(145,30){\makebox(0,0){$\rm NH_3D^+$}}

\put(178,50){\makebox(0,0){$\rm NH_3$}}
\put(178,30){\makebox(0,0){$\rm NH_2D$}}

\put(118,42){\vector(3,1){21}}
\put(118,38){\vector(3,-1){21}}
\put(128,48){\makebox(0,0){$\frac{1}{3}$}}
\put(128,31){\makebox(0,0){$\frac{2}{3}$}}

\put(152,50){\vector(1,0){18}}
\put(152,30){\vector(1,0){18}}
\put(152,30){\vector(4,3){18}}
\put(161,53){\makebox(0,0){$\frac{1}{1}$}}
\put(161,25){\makebox(0,0){$\frac{3}{4}$}}
\put(159,40){\makebox(0,0){$\frac{1}{4}$}}

\put(92,15){\line(0,1){55}}

\end{picture}

\caption{Branching ratio diagrams for the $\rm NH_3 + D_2H^+$ reaction in the FS ({\sl left}) and PH ({\sl right}) models, showing the re-distribution of hydrogen and deuterium in the reaction followed by dissociative recombination with electrons.}
\label{fig:PHvsFS}
\end{figure*}

Figure \ref{fig:PHvsFS} illustrates the difference between the models. Here we show the branching ratios arising from the PH and FS assumptions, when the $\rm NH_3 + D_2H^+$ reaction first forms $\rm NH_4^+$ (or one of its isotopologs), which subsequently dissociates mainly into $\rm NH_3 + H$ \citep{Ojekull04} in a dissociative recombination reaction with an electron. The PH reaction scheme is straightforward. The FS mechanism, however, can result in the formation of $\rm NH_2D_2^+$, which is not possible in the PH mechanism, and $\rm NH_2D_2^+$ can then decay into $\rm NHD_2$ which is not produced either in the PH scheme in these particular reactions. The FS model strongly favors the production of multiply deuterated species.

The PH mechanism as described above can be applied to reactions where one proton is transferred between the reactants, such as proton-donation reactions of the form $\rm XH^+ + Y \longrightarrow X + YH^+$, or hydrogen abstraction reactions such as the main ammonia formation reactions (\ref{eq:ammonia1}) to (\ref{eq:ammonia4}). Fiducially, we applied the PH mechanism to ion-molecule proton-donation reactions only, but we also considered a case where abstraction reactions are modified in similar fashion (see Sect.\,\ref{ss:HydrogenAbstraction}). For nearly all other reactions we assume that FS is valid, where applicable. A notable exception is charge-transfer reactions, where the H and D atoms cannot be mixed during the reaction.

We note that the approach described above to calculate the branching ratios in reactions involving deuterium is extremely simplistic, and is based purely on statistical considerations. It neglects, for example, any zero-point energy effects that come into play when introducing different counts of hydrogen or deuterium in the reactions. Detailed experimental and/or theoretical data is available only for a limited set of reacting systems including deuterated species. An example of this is the study of the $\rm H_3^+ + H_2$ system by \citet{Hugo09}, whose data we use here as well.

\subsection{FS versus PH: spin states}\label{ss:ScramblingVersusHopSpins}

Assuming PH instead of FS also changes the way that spin-state chemistry is treated in proton-donation reactions. This point has been discussed for example by \citet{Oka04} in the context of hydrogen. Here we extend the discussion to deuterium. Consider as an example the deuteron-donation reaction
\begin{equation}\label{eq:spinExample1}
\rm NH_2D_2^+ + D_2 \longrightarrow NH_2D + D_3^+ \, .
\end{equation}
We note that this reaction is endothermic and hence inefficient at low temperature. It was selected to illustrate the effect of the adopted reaction mechanism on product spin states when multiple deuterons are present in both reactants. In the FS scenario, the reaction proceeds through a complex that contains in this case two protons and four deuterons. The nuclear spin symmetries of protons ($I = \frac{1}{2}$) and deuterons ($I = 1$) must be treated separately. The treatment of $\rm H_2$ spin is trivial: the spin state is unaltered in this reaction. For deuterium we consider the subreaction\footnote{In this paper we refer to any systems that represent a part of a complete reaction, such as reaction~(\ref{eq:spinExample1}), as ``subreactions''.} $\rm D_2^+ + D_2 \longrightarrow \left( D_4^+ \right)^* \longrightarrow D + D_3^+$. The reactions between different spin states in this subreaction can be calculated using the group-theoretical method presented in detail in \citet{Sipila15b}. In brief, the governing principle is the conservation of nuclear spin, and the task is to determine the symmetry representations of the reactants, products, and the intermediate complex, and to derive correlation tables between the group of the complex and its subgroups which represent the reactants and products. The correlation tables yield all of the necessary information, i.e., which product spin species are possible given a particular pair of reactants. Example tables are given in \citet{Sipila15b} and we omit them here for brevity; see also \citet{Hily-Blant18} for systems not covered in \citet{Sipila15b}.

\begin{figure*}[hbt]
\unitlength=1.0mm
\begin{picture}(160,80)(0,-10)

\color{black} 


\put(45,60){\makebox(0,0){$\sl full~scrambling$}}

\put(20,40){\makebox(0,0){$\rm NH_2oD_2^+ + oD_2$}}

\put(65,50){\makebox(0,0){$\rm NH_2D + mD_3^+$}}
\put(65,30){\makebox(0,0){$\rm NH_2D + oD_3^+$}}

\put(32,42){\vector(3,1){21}}
\put(32,38){\vector(3,-1){21}}
\put(40,48){\makebox(0,0){$\frac{10}{18}$}}
\put(40,31){\makebox(0,0){$\frac{8}{18}$}}

\put(20,10){\makebox(0,0){$\rm NH_2oD_2^+ + pD_2$}}

\put(65,20){\makebox(0,0){$\rm NH_2D + mD_3^+$}}
\put(65,10){\makebox(0,0){$\rm NH_2D + pD_3^+$}}
\put(65,0){\makebox(0,0){$\rm NH_2D + oD_3^+$}}

\put(32,12){\vector(3,1){21}}
\put(32,10){\vector(1,0){21}}
\put(32,8){\vector(3,-1){21}}
\put(40,18){\makebox(0,0){$\frac{5}{18}$}}
\put(46,13){\makebox(0,0){$\frac{1}{18}$}}
\put(40,1){\makebox(0,0){$\frac{12}{18}$}}


\put(135,60){\makebox(0,0){{$\sl proton~hop$}}}

\put(115,40){\makebox(0,0){$\rm NH_2oD_2^+ + oD_2$}}

\put(160,50){\makebox(0,0){$\rm NH_2D + mD_3^+$}}
\put(160,30){\makebox(0,0){$\rm NH_2D + oD_3^+$}}

\put(127,42){\vector(3,1){21}}
\put(127,38){\vector(3,-1){21}}
\put(135,48){\makebox(0,0){$\frac{10}{18}$}}
\put(135,31){\makebox(0,0){$\frac{8}{18}$}}

\put(115,10){\makebox(0,0){$\rm NH_2oD_2^+ + pD_2$}}

\put(160,20){\makebox(0,0){$\rm NH_2D + pD_3^+$}}
\put(160,0){\makebox(0,0){$\rm NH_2D + oD_3^+$}}

\put(127,12){\vector(3,1){21}}
\put(127,8){\vector(3,-1){21}}
\put(135,18){\makebox(0,0){$\frac{1}{9}$}}
\put(135,1){\makebox(0,0){$\frac{8}{9}$}}

\put(92,0){\line(0,1){55}}

\end{picture}

\caption{Spin-state branching ratio diagrams for the $\rm NH_2oD_2^+ + (o/p)D_2$ reaction in the FS ({\sl left}) and PH ({\sl right}) models, omitting the $\rm H_2$ spin state.}
\label{fig:PHvsFS_spinStates}
\end{figure*}
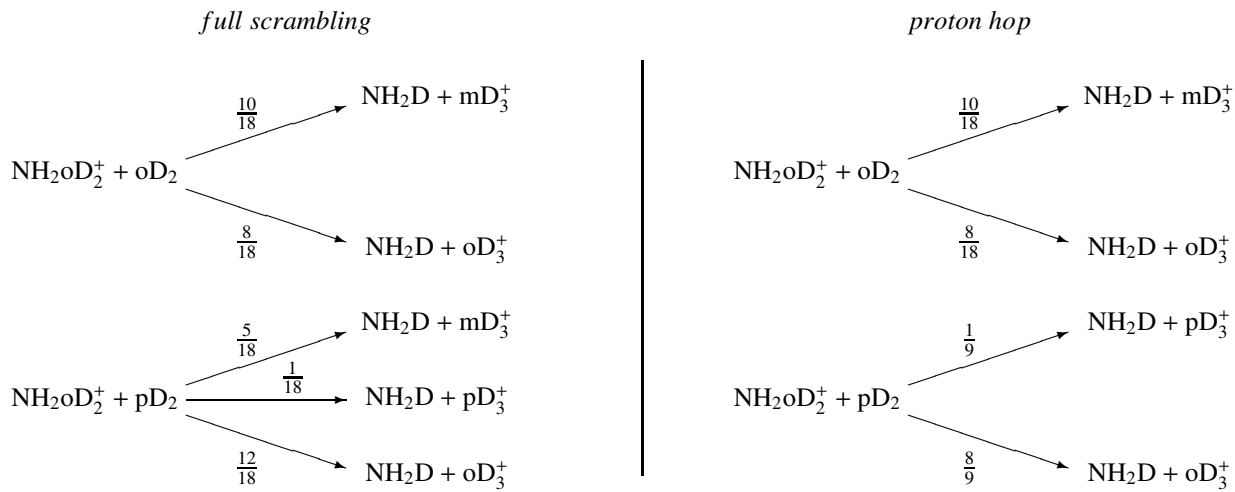

In the PH scenario, reaction~(\ref{eq:spinExample1}) is divided into two subreactions that take place in succession. First, the deuteron-donating ion ($\rm D_2^+$) breaks up:
\begin{equation*}
\rm D_2^+ \longrightarrow D^+ + D \, .
\end{equation*}
In the second step, the deuterium nucleus reacts with the neutral reactant:
\begin{equation*}
\rm D^+ + D_2 \longrightarrow D_3^+ \, .
\end{equation*}
Therefore reaction~(\ref{eq:spinExample1}) can be formally broken down to
\begin{equation}
\rm NH_2{\color{blue} DD^+} + {\color{red} D_2} \longrightarrow NH_2{\color{blue} D} + {\color{blue} D^+} + {\color{red} D_2} \longrightarrow NH_2{\color{blue} D} + {\color{red} D_2}{\color{blue} D^+} \, ,
\end{equation}
where the colors are added to help track the placements of the individual D atoms in the reaction. The branching ratios for each of the two subreactions are calculated using the method outlined above, i.e., full scrambling applies for the subreactions separately, but not for the reaction as a whole since it proceeds in two distinct steps. The resulting branches for reaction~(\ref{eq:spinExample1}) are shown in Fig.\,\ref{fig:PHvsFS_spinStates}, omitting the $\rm H_2$ spin state. We note that the reaction channel that forms p$\rm D_3^+$ is forbidden by selection rules when the neutral reactant is o$\rm D_2$. There are again fewer reaction pathways in the PH model than in the FS model. For the reaction involving o$\rm D_2$, the two schemes yield the same product channels with identical branching ratios, but when p$\rm D_2$ is involved, the formation of m$\rm D_3^+$ is forbidden in the PH model.

Ammonia formation is governed mainly by hydrogen atom abstraction reactions of the type ${\rm NH}_i^+ + {\rm H_2} \longrightarrow {\rm NH}_{i+1}^+ + {\rm H}$, such as the $\rm N^+ + H_2 \longrightarrow NH^+ + H$ reaction, and it is clear that considering these reactions as abstraction reactions -- as opposed to assuming them to proceed through scrambling as is our standard approach -- would have an effect on our results. We discuss this topic in detail in Sect.\,\ref{ss:HydrogenAbstraction}.

\subsection{Chemical model}\label{ss:ChemicalModel}

We use the gas-grain chemical code described in \citet{Sipila19}. In short, the code uses the rate-equation approach to calculate chemical time-evolution in the gas phase and on grain surfaces. Gas-phase and grain-surface chemistry are connected via adsorption and (non-thermal) desorption processes. We consider a two-phase model where the ice on dust grains consists of a single reactive bulk, i.e., no separation between a chemically active layer and an inert mantle is made.

As in our recent modeling papers \citep[e.g.,][]{Sipila19}, we took the latest public release of the KIDA gas-phase network (kida.uva.2014; \citealt{Wakelam15}) as our basis network, which was deuterated as described in Sect.\,\ref{ss:scramblingVersusHopDeuteration}. Spin-state chemistry was then introduced in the deuterated network following either the FS or PH methods discussed above. As in \citet{Sipila19}, we explicitly track the spin states of $\rm H_2$, $\rm H_2^+$, $\rm H_3^+$, and their deuterated isotopologs, as well as the spin states of any multiply protonated/deuterated species involved in the water and ammonia formation networks. We introduced deuteration and spin states in our grain-surface network \citep{Sipila19} in an analogous way, although we assume that no ions are present on or in the ice, and consider only the FS mechanism for those surface reactions where atom interchanges occur. The rate coefficients of electron recombination reactions involving the protonated ions in the main ammonia formation pathway are taken from the KIDA database. The products of the dissociative pathways follow statistical ratios. 

The final number of gas-phase reactions depends on whether we adopt the FS or the PH model. In the former case, the network contains $\sim$75000 reactions, while in the latter case the number is reduced to $\sim$64000. The number of grain-surface reactions is similar in both cases: $\sim$2400 versus $\sim$2200 in the FS and PH models, respectively.

\subsection{Physical models}

In this paper we demonstrate the differences arising from using either the FS or PH mechanism on the D/H and spin-state abundance ratios of various chemical species. We use zero-dimensional (0D) physical models for this purpose, i.e., we calculate the chemical evolution as a function of time in point-like fixed physical conditions. We adopt in the 0D models a constant temperature of $T_{\rm dust} = T_{\rm gas} = 10 \, \rm K$, a visual extinction of $A_{\rm V} = 10 \, \rm mag$, and a primary cosmic ray ionization rate per hydrogen atom of $\zeta_p({\rm H}) = 1.3 \times 10^{-17}\,\rm s^{-1}$.

\begin{table}
	\centering
	\caption{Initial abundances (with respect to the total proton number density $n_{\rm H} \approx 2 \times n({\rm H_2})$) used in the chemical modeling.}
	\begin{tabular}{l|l}
		\hline
		\hline
		Species & Abundance\\
		\hline
		$\rm H_2$ & $5.00\times10^{-1}$\\
		$\rm He$ & $9.00\times10^{-2}$\\
		$\rm HD$ & $1.60\times10^{-5}$\\
		$\rm C^+$ & $7.30\times10^{-5}$\\
		$\rm N$ & $5.30\times10^{-5}$\\
		$\rm O$ & $1.76\times10^{-4}$\\
		$\rm S^+$ & $8.00\times10^{-8}$\\
		$\rm Si^+$ & $8.00\times10^{-9}$\\
		$\rm Na^+$ & $2.00\times10^{-9}$\\
		$\rm Mg^+$ & $7.00\times10^{-9}$\\
		$\rm Fe^+$ & $3.00\times10^{-9}$\\
		$\rm P^+$ & $2.00\times10^{-10}$\\
		$\rm Cl^+$ & $1.00\times10^{-9}$\\
		\hline
	\end{tabular}
	\label{tab:initialabundances}
	\tablefoot{The initial $\rm H_2$ o/p ratio is $1.0 \times 10^{-3}$.}
\end{table}

\begin{figure}
\centering
\includegraphics[width=1.0\columnwidth]{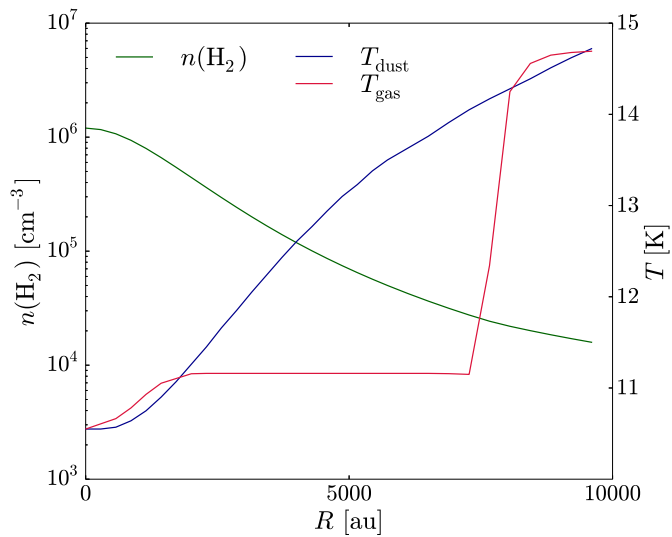}
\caption{Density and temperature profiles of the H-MM1 core model.
}
\label{fig:HMM1_physicalmodel}
\end{figure}

We also reproduce some results presented in H17, where we attempted to fit observations of deuterated ammonia in the starless core H-MM1 using a previous version of our chemical model. The model results presented therein were derived using the FS assumption. Here we compare the FS and PH mechanisms to see which one better matches the observations. We therefore used the core model derived in H17 and divided it into concentric spherical shells in which we calculated the chemical evolution. We then combined the results to produce radius-dependent abundance profiles, and compared them to the observed tendencies. The density and temperature profiles of the model core are displayed in Fig.\,\ref{fig:HMM1_physicalmodel}. We refer the reader to H17 for details on how these profiles were derived, but highlight here two prominent features of the gas temperature profile: 1) The profile is based on $\rm NH_3$(1,1) and (2,2) observations. The sharp increase in the temperature at around 7500 au is accompanied with a transition from subsonic to supersonic turbulence regime; 2) The gas temperature lies below the dust temperature outside the inner region where gas-dust coupling is important, which is probably due to efficient line cooling \citep[see for example][]{Sipila18}. The initial abundances, reproduced in Table~\ref{tab:initialabundances}, correspond to the EA1 set of \citet{Wakelam08}, except that the elemental nitrogen abundance was multiplied by a factor of 2.5 (H17). We do not consider chemical desorption because it leads in our two-phase model to ammonia abundances that are much higher than observed \citep{Sipila19}. The effect of modifying the initial $\rm H_2$ o/p ratio is discussed in Sect.\,\ref{ss:h-mm1}.

We also carried out radiative transfer simulations of the ammonia lines in order to compare our chemical modeling results with the observations of H17 in a meaningful way. These simulations are described in Sect.\,\ref{ss:LineSimulations}.

\section{Results}\label{s:results}

\subsection{0D models}\label{ss:0dModels}

\begin{figure*}
\includegraphics[width=2.0\columnwidth]{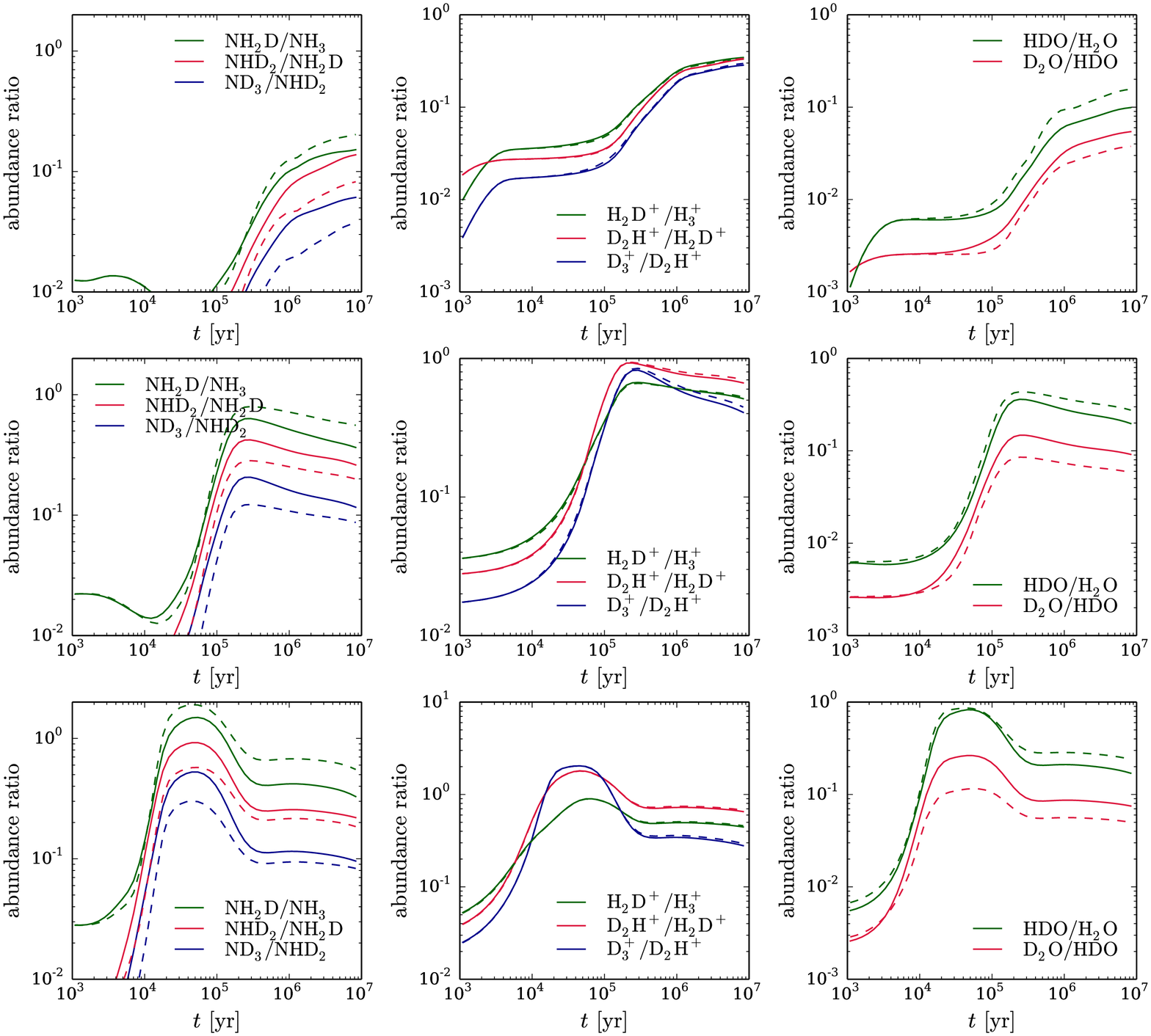}
\caption{D/H abundance ratios of ammonia ({\sl left column}), $\rm H_3^+$ ({\sl middle column}), and water ({\sl right column}), at a medium density of $n({\rm H_2}) = 10^4 \, \rm cm^{-3}$ ({\sl top row}), $n({\rm H_2}) = 10^5 \, \rm cm^{-3}$ ({\sl middle row}), and $n({\rm H_2}) = 10^6 \, \rm cm^{-3}$ ({\sl bottom row}). The abundances represent sums over the spin states where applicable. Solid lines represent the FS model, while dashed lines represent the PH model.
}
\label{fig:FSvsPH_abus}
\end{figure*}

\begin{figure*}
\includegraphics[width=2.0\columnwidth]{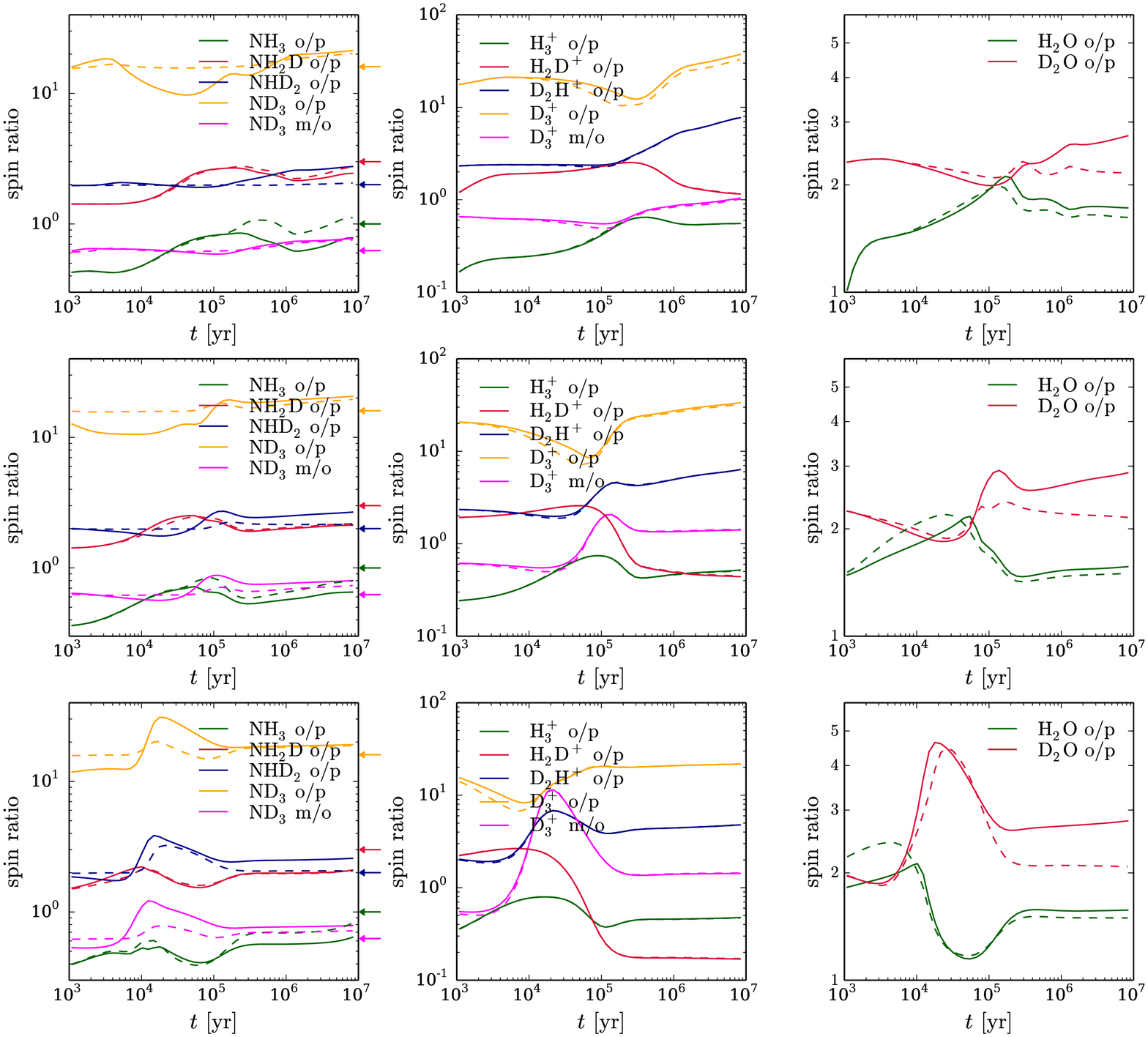}
\caption{Spin-state abundance ratios of ammonia ({\sl left column}), $\rm H_3^+$ ({\sl middle column}), and water ({\sl right column}), including their respective deuterated forms, at a medium density of $n({\rm H_2}) = 10^4 \, \rm cm^{-3}$ ({\sl top row}), $n({\rm H_2}) = 10^5 \, \rm cm^{-3}$ ({\sl middle row}), and $n({\rm H_2}) = 10^6 \, \rm cm^{-3}$ ({\sl bottom row}). Solid lines represent the FS model, while dashed lines represent the PH model. The statistical ammonia spin-state abundance ratios are indicated with colored arrows.
}
\label{fig:FSvsPH_ratios}
\end{figure*}

Figures~\ref{fig:FSvsPH_abus} and \ref{fig:FSvsPH_ratios} show the time-evolution of various D/H and spin-state ratios at different medium densities in the FS and PH models at $T = 10\,\rm K$. First, it is evident that $\rm H_3^+$ and its isotopologs are practically unaffected by the branching mechanism. This is because the chemistry of these species is mainly determined by the $\rm H_3^+ + H_2$ reacting system, for which we adopt the rate coefficients calculated by \citet{Hugo09} regardless of what we assume for any other proton-donation reactions. The feedback of reactions such as those illustrated in Figs.\,\ref{fig:PHvsFS}~and~\ref{fig:PHvsFS_spinStates} on (deuterated) $\rm H_3^+$ is evidently very small.

Both the D/H and spin ratios of ammonia are in turn clearly affected by the branching mechanism, although the spin ratios experience a lesser effect. Adopting the PH mechanism increases the $\rm NH_2D / NH_3$ ratio, but decreases the $\rm NHD_2 / NH_2D$ and $\rm ND_3 / NHD_2$ ratios by about the same factor. While we recover similar time-dependence in the D/H ratios as in our previous work \citep{Sipila15b}, we obtain generally higher values for the ratios, and notably the $\rm NH_2D / NH_3$ ratio can exceed unity in the current model. There is an evident peak for the ratios which depends on time; this occurs at $\sim2 \times 10^5 \, \rm yr$ at $n({\rm H_2}) = 10^5 \, \rm cm^{-3}$, for example. The difference between the FS and PH models increases at late times for the $\rm NH_2D / NH_3$ ratio, but decreases for $\rm NHD_2 / NH_2D$ and $\rm ND_3 / NHD_2$. The ammonia spin ratios show some density dependence in the difference between the FS and PH models, although this is always a factor of two at most. The $\rm ND_3$ spin ratios are affected the most when changing the model. This is expected, given that $\rm ND_3$ has three distinct spin states, and the PH model includes significantly less reaction channels for $\rm ND_3$ formation. Statistical ammonia D/H ratios obey $\rm (NH_2D / NH_3) / (NHD_2 / NH_2D) = 3$ and $\rm (NHD_2 / NH_2D) / (ND_3 / NHD_2) = 3$ (\citealt{Brown89}; H17). The PH model predicts for these ratios (at late times at high density) values of $\sim$2.9 and $\sim$2.2, respectively. The corresponding values in the FS model are $\sim$1.45 and $\sim$2.3. We discuss these ratios in more detail in Sect.\,\ref{ss:HydrogenAbstraction}.

The water D/H ratios present similar behavior to the ammonia ones: the PH model predicts more deuteration in singly-substituted molecules, but less deuteration in multiply-substituted molecules. The difference in the HDO/$\rm H_2O$ and $\rm D_2O$/HDO ratios between the two models is almost density-independent. The $\rm H_2O$ o/p ratio is largely unaffected by the choice of branching mechanism. The $\rm D_2O$ o/p ratio presents small deviations (of the order of a few 10\%; note the y-axis scale in Fig.\,\ref{fig:FSvsPH_ratios}) depending on the model, which are attributed to the spin ratios of its precursors $\rm HD_2O^+$ and $\rm D_3O^+$ (in particular the most abundant meta (m) and ortho forms). For example, m$\rm D_3O^+ + e^-$ can only lead to o$\rm D_2O$ formation while o$\rm D_3O^+ + e^-$ can form ortho and para $\rm D_2O$ with equal probability. At high density in the FS model, the $\rm D_3O^+$ m/o ratio is above unity while in the PH model it is below unity, meaning that production of para-$\rm D_2O$ is more efficient in the PH model, decreasing the $\rm D_2O$ o/p ratio with respect to the FS model.

The general conclusion from these results is that the PH model favors singly-deuterated species over multiply-deuterated ones and produces deuterated species slightly more efficiently in the higher-energy spin states, that is, $\rm pNH_2D$, $\rm pNHD_2$, and
$\rm oND_3$, when compared to the FS model. (The energy level diagrams of the various ammonia isotopologs are shown in Fig.\,3 in H17.) The former conclusion was already expected based on our discussion in Sect.\,\ref{s:model}, where the example reaction showed the tendency of the PH model to omit multiply-deuterated pathways.

\subsection{H-MM1}\label{ss:h-mm1}

\begin{figure*}
\centering
\includegraphics[width=2.0\columnwidth]{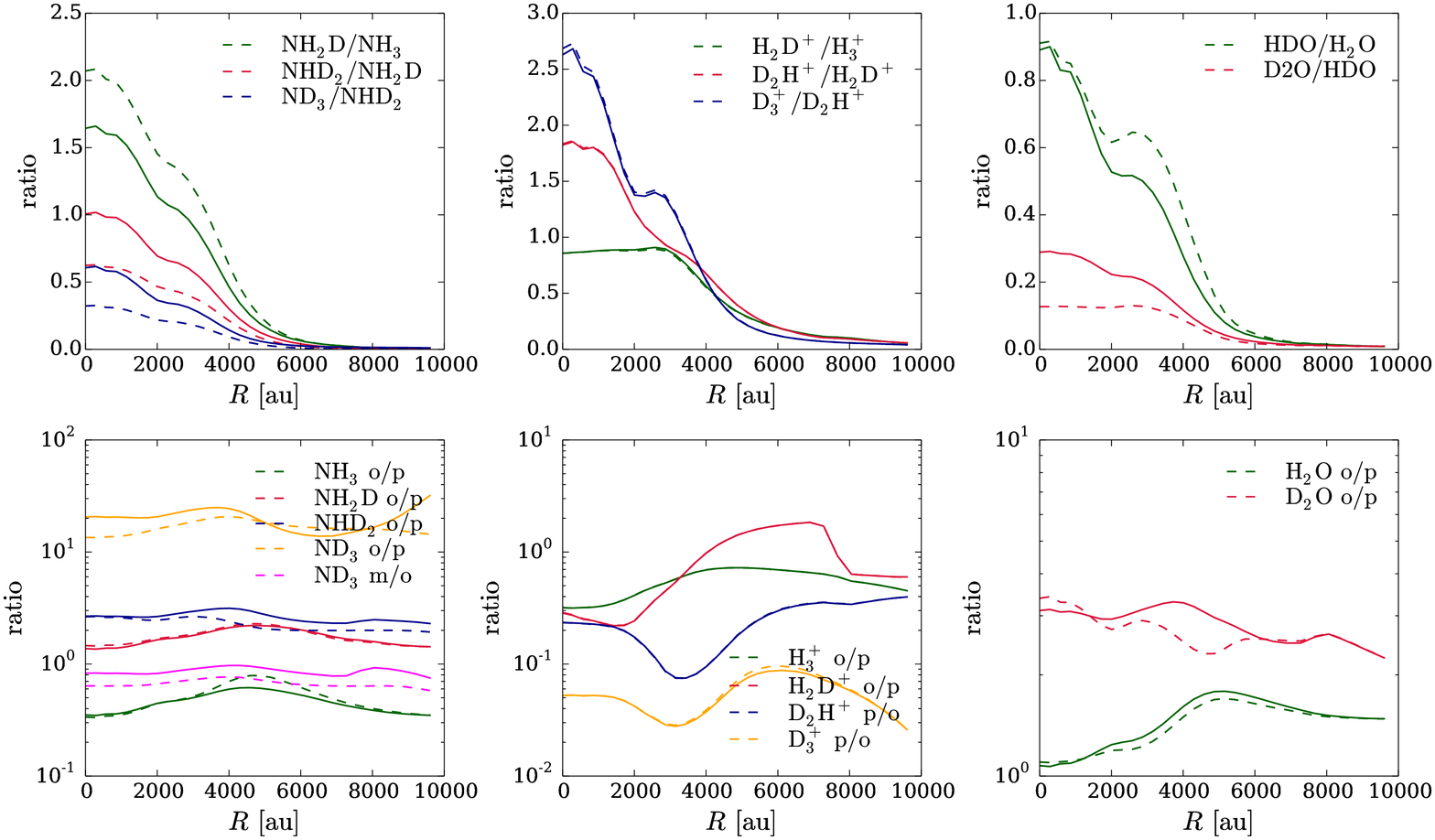}
\caption{D/H ratios ({\sl top row}) of ammonia ({\sl left column}), $\rm H_3^+$ ({\sl middle column}), and water ({\sl right column}) and the corresponding spin-state ratios ({\sl bottom row}), as functions of radius in the H-MM1 core model. The curves correspond to $t = 10^5 \, \rm yr$. Solid lines represent the FS model, while dashed lines represent the PH model.
}
\label{fig:H-MM1}
\end{figure*}

Figure~\ref{fig:H-MM1} shows the D/H and spin ratios of the same species as in Figs.\,\ref{fig:FSvsPH_abus}~and~\ref{fig:FSvsPH_ratios}, but as radial profiles calculated using the H-MM1 core model. The results correspond to $t = 10^5 \, \rm yr$ since the start of the simulation. (The time was arbitrarily chosen for the sake of illustration.) The same trends already discussed above are present here as well: for example, the $\rm NHD_2/NH_2D$ and $\rm ND_3/NHD_2$ ratios are clearly lower in the core center in the PH model. The sharp decrease in the $\rm H_2D^+$ o/p ratio near the edge of the core is due to the sudden increase in the gas temperature (Fig.\,\ref{fig:HMM1_physicalmodel}). A corresponding feature is present in some of the other spin ratios as well, although it is far less prominent for the other ratios.

\begin{figure*}
\centering
\includegraphics[width=2.0\columnwidth]{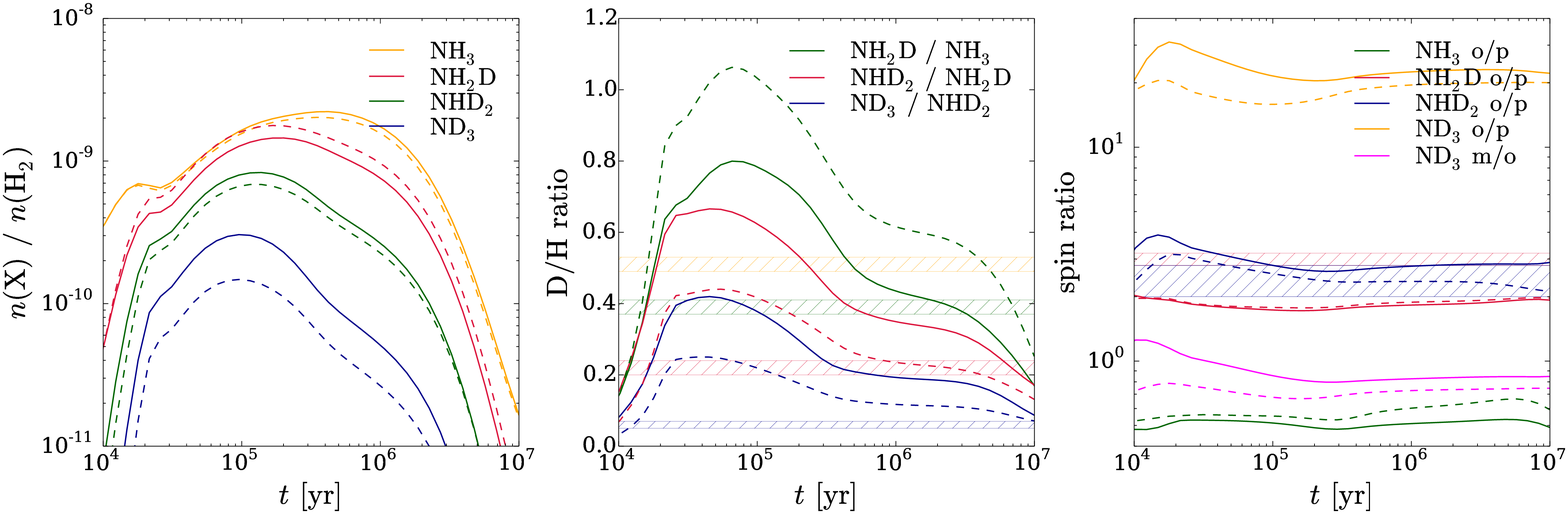}
\caption{Abundances of the ammonia isotopologs summed over the spin states ({\sl left}), corresponding D/H ratios ({\sl middle}), and spin-state ratios ({\sl right}), as functions of time in the H-MM1 core model. Solid lines represent the FS model, while dashed lines represent the PH model. Observational ratios derived by H17 are overlaid ({\sl hatched boxes}). The orange hatched box in the middle panel represents the $\rm NH_2D/NH_3$ ratio calculated using an $\rm NH_3$ o/p ratio of 0.5 (see text for details).
}
\label{fig:H-MM1averages}
\end{figure*}

The D/H and spin ratios are heavily time-dependent, and it is difficult to draw any firm conclusions on the difference between the model and observations based on the radial profiles presented in Fig.\,\ref{fig:H-MM1}. To complement that figure, we show in Fig.\,\ref{fig:H-MM1averages} the density-weighted abundances and abundance ratios of the ammonia isotopologs as functions of time. The observed D/H and spin ratios derived by H17 are overlaid. The early-time behavior of the various ammonia abundances is in the present model very different from that of H17. This is because of several updates to the chemical model since \citet{Sipila15b} that affect the early-time (due to ion-molecule reactions) and late-time (due to cosmic rays) chemistry in particular \citep[see][]{Sipila19}. However, we recover similar abundances compared to H17 in the time interval $10^5$ to $10^6$\,yr; the current model predicts a factor of a few more $\rm ND_3$.

The D/H ratios again follow the same trends as in the discussion above -- for example the $\rm NH_2D/NH_3$ ratio is enhanced in the PH model. Notably, the FS model can only fit the $\rm NH_2D/NH_3$ ratio for realistic timescales ($\gtrsim 10^5 \, \rm yr$) where the abundances are not below detectable levels. The PH model reproduces the observed $\rm NHD_2/NH_2D$ ratio within the observational uncertainties, and simultaneously fits the observed $\rm ND_3/NHD_2$ ratio within a factor of two. While the $\rm NH_2D/NH_3$ ratio is better reproduced by the FS model, we note that H17 only detected para-$\rm NH_3$ and derived the $\rm NH_2D/NH_3$ ratio using an $\rm NH_3$ o/p ratio of unity. If we instead assume $\rm NH_3$ o/p = 0.5 as suggested by our chemical model, the PH model can match the observed ratio (orange hatched box) fairly well.

To summarize, the PH model can reproduce all three D/H ratios at a similar timescale, whereas the FS model can only reproduce the $\rm NH_2D/NH_3$ ratio. We note that H17 derived the observed abundance ratios assuming constant abundances throughout the core, and this is in stark contrast with the chemical model which predicts strong abundance gradients. Indeed, neither the FS or the PH model can match the observed lines well, as we show in Sect.\,\ref{ss:LineSimulations}.

The observed $\rm NHD_2$ o/p ratio is reproduced by both models, and both underestimate the $\rm NH_2D$ o/p ratio by about 50\%. The similarities in the spin ratios of the ammonia isotopologs in the two models is unsurprising given that the ratios are dominated by reactions with $\rm H_3^+$ isotopologs \citep[see the discussion in][]{Sipila15b}, and in this paper we use the same chemistry for the $\rm H_3^+ + H_2$ system \citep{Hugo09} in both the FS and PH cases. Hence a modification to the $\rm H_3^+ + H_2$ system would be required to obtain a better agreement with the observed ammonia spin state ratios. This is beyond the scope of the present work; see Sects.\,\ref{ss:SpeciesToSpecies}~and~\ref{ss:HydrogenAbstraction} for some additional discussion.

We adopted in our models a low initial $\rm H_2$ o/p ratio of $10^{-3}$, assuming that the ratio has been thermalized to 20\,K before the dense core stage (H17). Low-temperature deuterium chemistry is very sensitive to the $\rm H_2$ o/p ratio, and hence it is clear that changing the initial value would influence our results, at least at early times in the simulation. To test this, we ran the FS and PH models assuming instead an initial $\rm H_2$ o/p ratio of 0.1. This modification leads to very low abundances of deuterated species at early times (before $\sim$$10^5\,\rm yr$) because the presence of ortho-$\rm H_2$ suppresses deuteration, and moves the abundance peaks displayed in Fig.\,\ref{fig:H-MM1averages} forward by a factor of $\sim$2. The results of these test models and those of our fiducial FS and PH models are very close to each other for $t \gtrsim \rm a \, few\, \times 10^5\,\rm yr$, and in fact at late times the test models present very slightly higher D/H ratios (of the order of a few per cent) for the various species than the corresponding fiducial models do.

\section{Discussion}\label{s:discussion}

\subsection{Variation in the physical core model}\label{ss:PhysicalModelVariation}

The results discussed above suggest that the PH model is better than the FS model in reproducing the ammonia D/H ratios observed toward H-MM1. This result holds within the assumption that our physical model for H-MM1 represents the actual core well. The core model is based on observationally-derived $\rm H_2$ column density and gas/dust temperature profiles (see H17 for details). The observational error in the $\rm H_2$ column density is less than a factor of two, but for the gas temperature the error ranges from less than one kelvin to several kelvin depending on the distance from the center of the core, which can affect the results of our modeling. An important question in the present context is: can the uncertainty in the core model affect the goodness of the fit in the PH model versus FS model, i.e., is it possible that for another physical model the PH model would no longer be the better match to the observations? Our 0D calculations show that the difference between the two models does not present strong density variations, but to further quantify this issue we ran additional core models, varying the density and gas temperature profiles. In these tests our fiducial density profile was scaled by a factor of $\frac{1}{2}$, 1, or 2, and the fiducial gas temperature profile was shifted by -1, 0, or 1\,K for a total of nine models including our fiducial model. We note that although the observational error in the gas temperature may be several kelvin, in the inner core where most of the line emission originates (depending on the species and the line) the error is of the order of one kelvin only (H17), and hence shifting the gas temperature by this conservative value is justified. We do not alter the dust temperature profile for the sake of simplicity.

\begin{figure*}
\centering
\includegraphics[width=2.0\columnwidth]{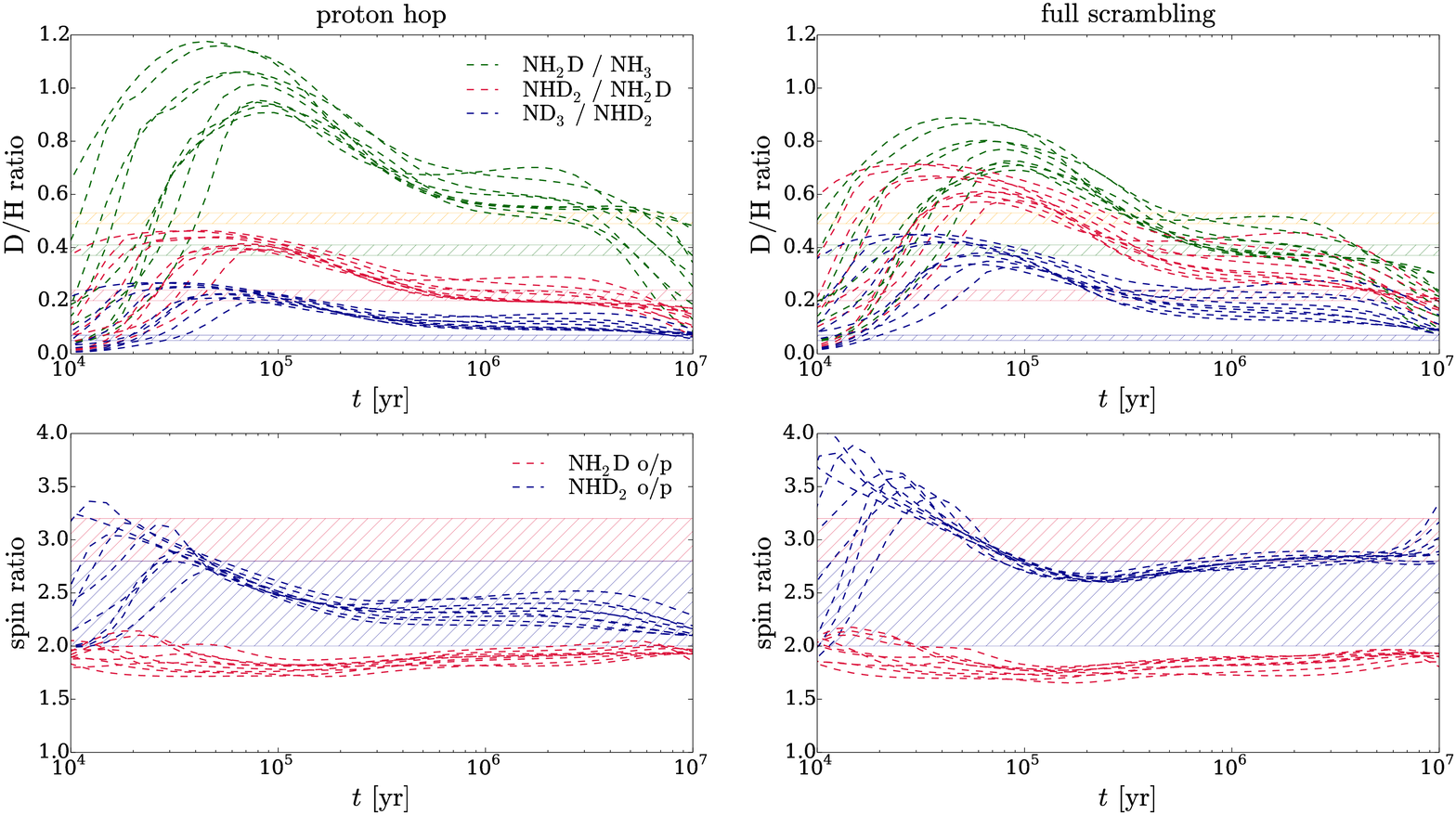}
\caption{{\sl Left:} D/H and spin-state abundance ratios in the PH model, calculated using nine different physical core models as explained in the text. {\sl Right:} Corresponding data from the FS model. Hatched boxes correspond to observed ranges as in Fig.\,\ref{fig:H-MM1averages}.
}
\label{fig:H-MM1averages_multiplemodels}
\end{figure*}

Figure~\ref{fig:H-MM1averages_multiplemodels} shows the results of these calculations, with all nine cases overlaid. There is some spread in the ratios owing to the changes in the physical conditions, but an analysis of the individual curves shows that the PH model always yields a higher $\rm NH_2D/NH_3$ ratio, and lower $\rm NHD_2/NH_2D$ and $\rm ND_3/NHD_2$ ratios, than the FS model does, no matter which parameter combination we use. This strengthens the view that the PH model is indeed the better match to the observed abundance ratios.

The observed $\rm NH_2D/NH_3$ ratio (assuming $\rm NH_3 \, \rm o/p = 0.5$) can be reached by the PH model if the density of the model core is multiplied by a factor of $\frac{1}{2}$, and the best match is obtained if the fiducial temperature profile is either unaltered or shifted down by one kelvin. The $\rm ND_3/NHD_2$ ratio is also best fit by these parameter combinations; in those cases the $\rm NHD_2/NH_2D$ ratio is just at the observed lower limit at $t = 10^6 \, \rm yr$. Regardless of the changes to the density and temperature, we never obtain an $\rm NH_2D$~o/p ratio higher than~$\sim$2 in our models, marking a clear discrepancy with the observations.

\subsection{Line emission profiles}\label{ss:LineSimulations}

The match between the models and observations is best quantified by simulating line emission and comparing it with the observed lines, so that possible optical depth and excitation effects can be taken into account. H17 presented a detailed analysis of ammonia in H-MM1 including the lines of several other species, such as $\rm H_2D^+$ and $\rm N_2D^+$, and searched for the best match between models and observations by tuning the chemical and physical model parameters (e.g., interstellar radiation field strength, diffusion-to-binding energy ratio on grain surfaces, etc.). In this paper we perform only a reduced version of their analysis, concentrating on the ammonia lines. We adopt the radiative transfer pipeline used by H17 and apply it to the results from the up-to-date chemical models presented above.

We extracted the abundance profiles of the eight distinct (deuterated) ammonia isotopologs (taking into account spin states) observed by H17 at 51 time steps spaced between $10^5$ and $10^7$ yr, and carried out line simulations for all of the eight species at each time step using the non-local-thermal-equilibrium Monte Carlo radiative transfer code of \citet{Juvela97}. We then searched for the best match between the observations and models by performing Pearson's $\chi^2$-test on the simulated lines\footnote{For the summation we only included channels for which the main beam temperature is greater than $3\sigma$, as determined from a line-free part of the observed spectrum.}
\begin{equation}
\chi^2 = \sum_i \frac{ \left( T_{\rm MB}^{\rm obs}(i) - T_{\rm MB}^{\rm mod}(i) \right)^2 }{T_{\rm MB}^{\rm mod}(i)} \, ,
\end{equation}
where the summation is over spectral channels. The final $\chi^2$ value at each time step is a simple mathematical average of seven species; we exclude para-$\rm ND_3$ from the test because it was not detected by H17. The best-fit time that we obtain is $t = 3.29 \times 10^5 \, \rm yr$ for both the FS and the PH model. The various modeled D/H ratios are closest to the observed ratios at around $t = 1 \times 10^6 \, \rm yr$, but at these late times the abundances of the deuterated ammonia isotopologs are too low (Fig.\,\ref{fig:H-MM1averages}) to match the intensities of the observed lines.

\begin{figure*}
\centering
\includegraphics[width=1.8\columnwidth]{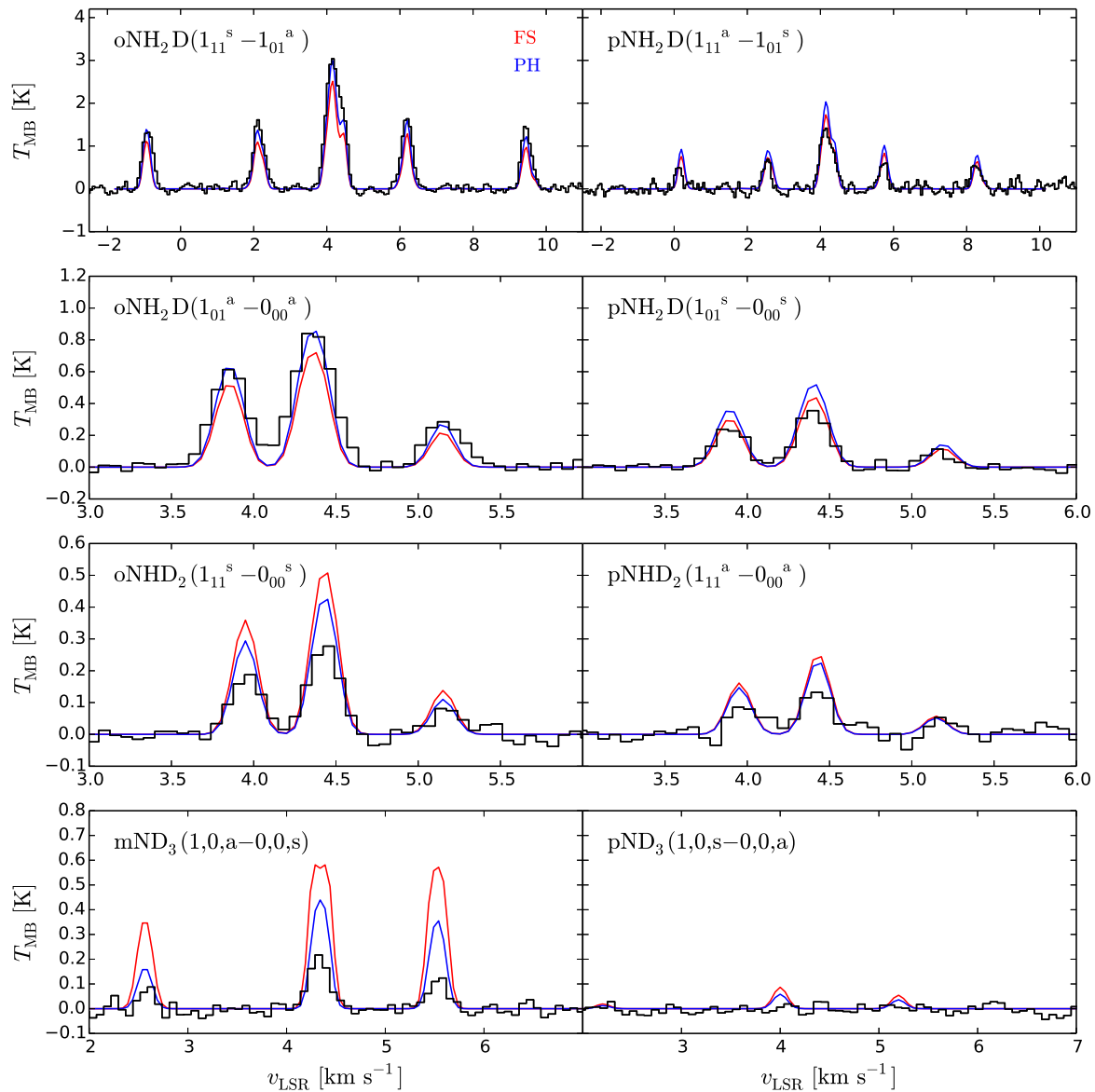}
\caption{Simulated emission lines for deuterated ammonia (specific species and line indicated in the panels) in the FS model (red) and PH model (blue). The black histograms show the observed lines (H17). The results correspond to the best-fit time $t = 3.29 \times 10^5 \, \rm yr$ (see text).
}
\label{fig:linesimulations}
\end{figure*}

Figure~\ref{fig:linesimulations} shows the best-fit simulated lines along with the lines observed by H17. We find that: 1) the para-$\rm NH_2D$ lines are overestimated by the model, which is unsuprising given that the model predicts a lower $\rm NH_2D$ o/p ratio than what is deduced from the observations; 2) The $\rm NHD_2$ and $\rm ND_3$ lines are clearly overestimated by the current chemical model, but the PH model reproduces the relative strength of the ortho and para lines slightly better; 3) The PH model is overall the better match to the observations, although the difference is not great. The line simulation results confirm the overall tendencies noted for the abundance distributions discussed above.

The agreement with the observations obtained by H17 using the chemical model of \citet{Sipila15b} is better than what we obtain with the present model. In this paper we did not attempt to modify the chemical model parameters for the purposes of obtaining the best possible match to the observations, and the present results do not imply that the new model performs worse. Indeed, if we explored the parameter space in the same way as H17 did, the match between the model and the observations could be improved. We did carry out another set of calculations using one of our alternative core models where the fiducial density profile is multiplied by $\frac{1}{2}$ and the gas temperature profile is shifted down by one kelvin. We found that in this case the overall fit is worse than in the regular FS or PH models, and already at $t = 3.29 \times 10^5 \, \rm yr$ the $\rm NH_2D$ and $\rm NHD_2$ lines are too weak. The results highlight the fact that line emission simulations are essential for comparing observations and chemical models.

\subsection{Effect of species-to-species rate coefficients}\label{ss:SpeciesToSpecies}

In \citet{Sipila17b} we presented a new set of rate coefficients for the $\rm H_3^+ + H_2$ reacting system, taking rotational excitation into account in the derivation of the rate coefficients (see \citealt{Hily-Blant18} for a recent similar work). We found that the inclusion of excited states leads in general to reduced deuteration levels. This effect is prominent for densities above $10^6 \, \rm cm^{-3}$ and temperatures above 10\,K, but we also found that the ortho/para ratio of $\rm H_2D^+$ is affected already at a density of a few times $10^5 \, \rm cm^{-3}$ because of the critical density of the ground-state $1_{10}$-$1_{11}$ rotational transition of o$\rm H_2D^+$. Given the dependency of the ammonia o/p (and D/H) ratios on the spin states of the $\rm H_3^+$ isotopologs, it is conceivable that rotational excitation could also modify the ammonia ratios indirectly. We tested this issue and found no significant effect on our results apart from a decrease in the $\rm H_2D^+$ o/p ratio of a few tens of per cent across the core. The density and temperature of H-MM1 are too low for rotational excitation of the $\rm H_3^+$ isotopologs to affect ammonia chemistry. This is in line with the discussion in \citet{Sipila17b} where we pointed out that the new rates are more important in protostellar disks than in starless or pre-stellar cores. However, we do obtain a noticeable effect on the $\rm H_2D^+$ o/p ratio which indicates that including rotational excitation is important for the correct interpretation of $\rm H_2D^+$ line emission in starless and pre-stellar cores, as already clearly demonstrated in \citet{Harju17b}.

\subsection{Abundances of additional species}

\begin{figure*}
\centering
\includegraphics[width=1.5\columnwidth]{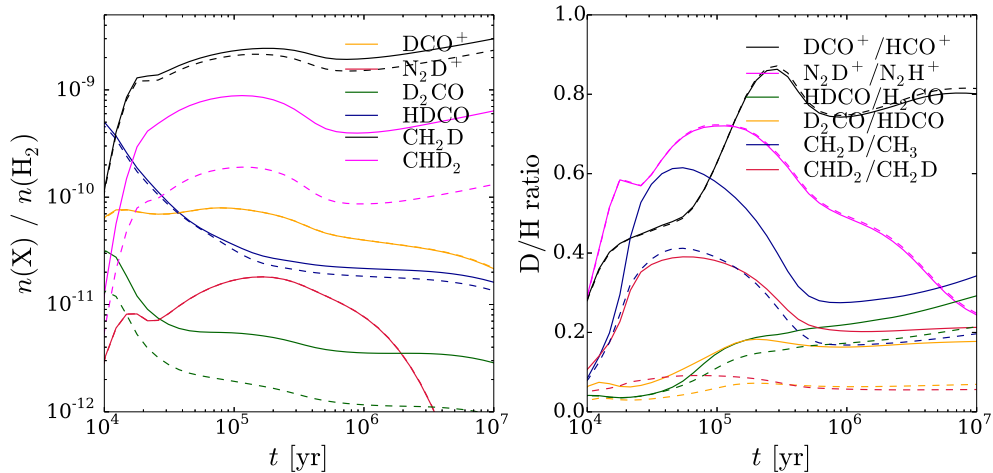}
\caption{As the left and middle panels of Fig.\,\ref{fig:H-MM1averages}, but showing the abundances and D/H ratios of some additional species.
}
\label{fig:H-MM1averages_additional}
\end{figure*}

We concentrated mainly on ammonia in the above analysis because of the direct comparison with our previous observations toward H-MM1. However, the nature of the proton-transfer mechanism has the potential to affect a wide variety of species other than ammonia, either directly or indirectly. In Fig.\,\ref{fig:H-MM1averages_additional}, we show the abundances and D/H ratios $\rm N_2D^+$, $\rm DCO^+$, HDCO, $\rm D_2CO$, $\rm CH_2D$, and $\rm CHD_2$. Of this list of species, $\rm N_2D^+$ and $\rm DCO^+$ are not affected by the choice of the model. This is not surprising given that they are both straightforward derivatives of $\rm H_2D^+$ which is itself virtually unaffected by the transfer mechanism as explained above. Deuterated formaldehyde shows behavior that is different from water for example: both the singly and doubly deuterated variants are produced less efficiently in the PH model. Formaldehyde is produced (at high density) in the dissociative electron recombination of $\rm CH_2OH^+$ with free electrons, and in neutral-neutral reactions of atomic oxygen with $\rm CH_3$. The fraction of deuterated $\rm CH_3$ is higher in the FS model than in the PH model -- this derives from differences in the chemistry of deuterated $\rm CH_4^+$ -- leading to more efficient production of HDCO and $\rm D_2CO$ through the neutral-neutral pathway. We note that (deuterated) $\rm H_2CO$ and $\rm CH_3$ have spin states that could in principle be tracked using our spin-separation routine, but we leave such an analysis for future work.

\subsection{Hydrogen atom abstraction reactions}\label{ss:HydrogenAbstraction}

The analysis presented above shows that neither the FS or the PH model can reproduce the $\rm NH_2D$ o/p abundance ratio observed toward H-MM1 regardless of variations in the physical core model, and that the various modeled D/H ratios also differ from the observed values. Although we did not perform an exhaustive parameter-space exploration, it seems clear that the observed ratios cannot be reproduced by modifying the reaction mechanism in proton-donation reactions only. The main ammonia formation reactions are abstraction reactions, and hence we constructed another chemical model where we assume that all ion-molecule hydrogen abstraction reactions analogous to reactions~(\ref{eq:ammonia1})~to~(\ref{eq:ammonia4}) proceed through PH instead of FS. The efficiencies of some of these reactions depend on the inclusion of deuterium and/or spin states. For example, reaction~(\ref{eq:ammonia1}) proceeds much faster with ortho-$\rm H_2$ than with para-$\rm H_2$ \citep{Dislaire12}, while the substitution of deuterium in $\rm H_2$ slows down the reaction \citep{Marquette88}. Reaction~(\ref{eq:ammonia4}) is slow at low temperature, although it does increase somewhat in efficiency toward 10\,K \citep{Barlow87}. The branching ratios for the modified abstraction reactions are calculated in a way analogous to what is described in Sect.\,\ref{s:model} for the proton-donation reactions. For example, the reaction $\rm NH_2D^+ + H_2$ can then only lead to $\rm NH_3D^+ + H$ -- the product pair $\rm NH_4^+ + D$ is not accessible like it is in the FS model, meaning that this alternative model (hereafter $\rm PH_{alt}$) is expected to produce larger deuterium fractions as many formation pathways leading to hydrogenated species are no longer available, compared to the FS model. Changing the reaction mechanism naturally also leads to repercussions in spin-state ratios. Proton-donation reactions and hydrogen abstraction reactions are both treated under the PH assumption in the $\rm PH_{alt}$ model.

\begin{figure*}
\centering
\includegraphics[width=2.0\columnwidth]{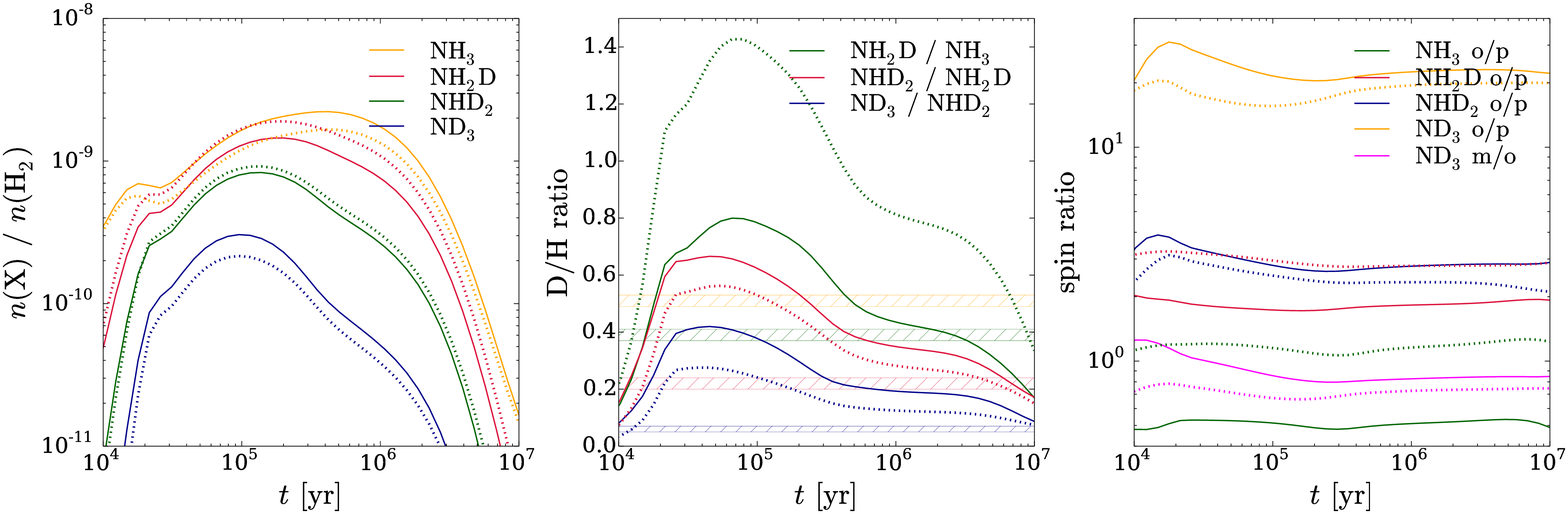}
\caption{As Fig.\,\ref{fig:H-MM1averages}, but showing the averaged abundances in H-MM1 in the FS model (solid lines) and in the $\rm PH_{alt}$ model discussed in the text (dotted lines). The hatched boxes showing the observed ranges in the right panel are omitted here for better visibility of the model results.
}
\label{fig:H-MM1averages_noabstraction}
\end{figure*}

Figure~\ref{fig:H-MM1averages_noabstraction} shows the averaged abundances in H-MM1 calculated with the FS model and the $\rm PH_{alt}$ model. Switching to a direct hydrogen atom hop mechanism for the abstraction reactions greatly increases the D/H ratios, as expected, and indeed the $\rm NH_2D / NH_3$ and $\rm NHD_2 / NH_2D$ ratios are now both higher than the observed ones. $\rm PH_{alt}$ produces ortho and para ammonia in roughly equal proportions, and in fact the modeled $\rm NH_2D / NH_3$ abundance ratio is now over a factor of two higher than the observed value if we account for the $\rm NH_3$ o/p ratio in the $\rm PH_{alt}$ model ($\sim$1.2). The value of the $\rm NH_2D$ o/p ratio ($\sim$2.7) at $t \sim 10^6 \, \rm yr$ is close to the observed value of $3.0 \pm 0.2$. Likewise, the o/p ratio of $\rm NHD_2$ approaches 2 at late times. This indicates that the modified abstraction reaction mechanism may give a better representation of ammonia spin-state ratios if the problem of excessively high D/H ratios can be countered.

\begin{figure*}
\centering
\includegraphics[width=1.8\columnwidth]{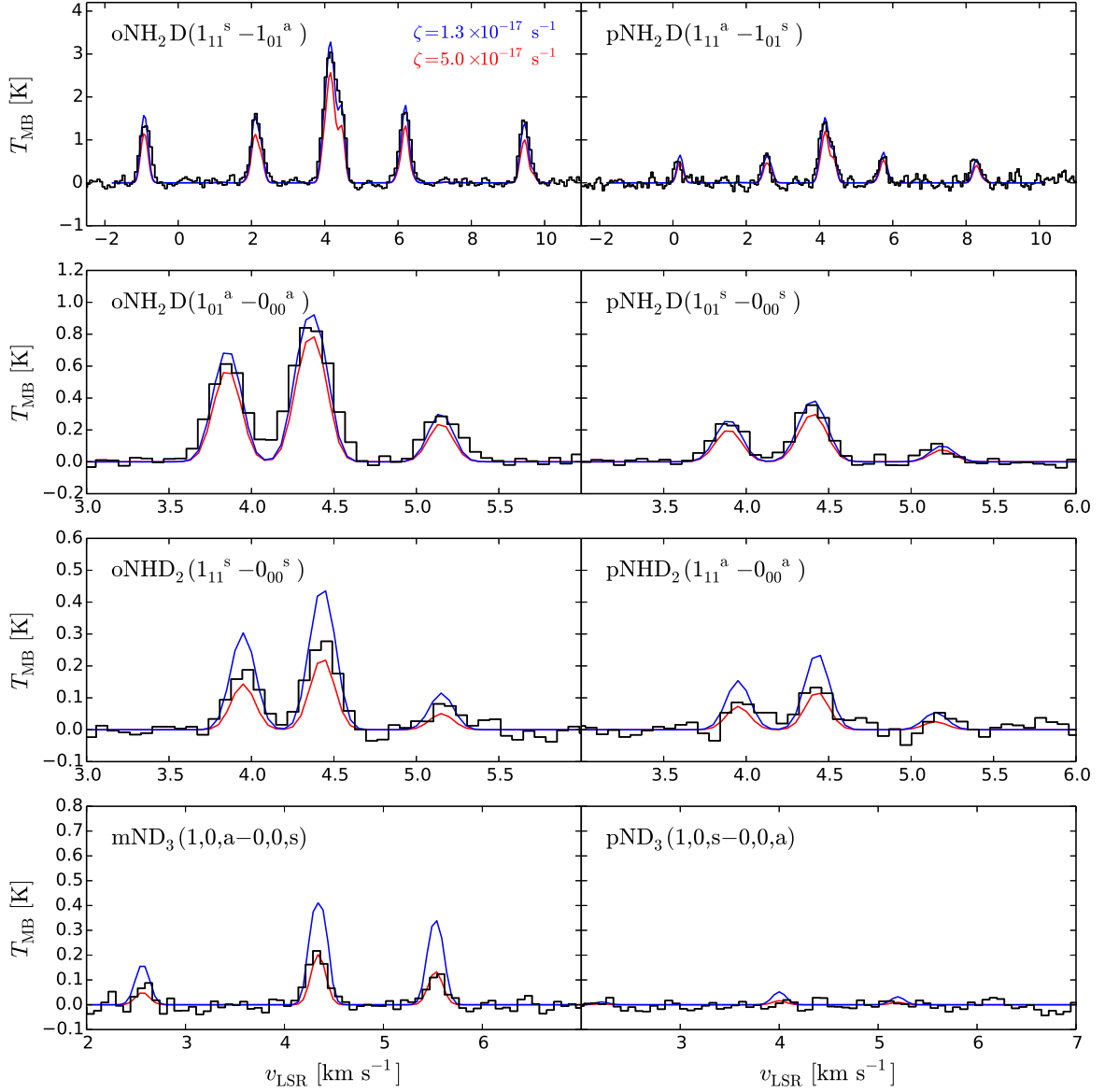}
\caption{As Fig.\,\ref{fig:linesimulations}, but for the $\rm PH_{alt}$ model assuming either $\zeta_p({\rm H}) = 1.3 \times 10^{-17}\,\rm s^{-1}$ (blue) or $\zeta_p({\rm H}) = 5.0 \times 10^{-17}\,\rm s^{-1}$ (red). The corresponding best-fit chemical times are $t = 4.33 \times 10^5 \, \rm yr$ and $t = 2.50 \times 10^5 \, \rm yr$.
}
\label{fig:linesimulations_noabstraction}
\end{figure*}

In \citet{Sipila10}, we found that increasing the cosmic-ray ionization rate leads to less efficient production of the deuterated $\rm H_3^+$ isotopologs, hence decreasing the efficiency of deuteration in general. Motivated by this result, we tested the influence of the cosmic-ray ionization rate on the H-MM1 model results by varying the ionization rate between $\zeta_p({\rm H}) = 1.0 \times 10^{-17}\,\rm s^{-1}$ and $1.0 \times 10^{-16}\,\rm s^{-1}$. We found that increasing $\zeta_p({\rm H})$ does indeed decrease the ammonia D/H ratios, but it also causes the abundances of the ammonia isotopologs to drop below detectable levels in an increasingly shorter timescale, of the order of a few times $10^5\,\rm yr$. However, the spin-state abundance ratios of the ammonia isotopologs are insensitive to changes in $\zeta_p({\rm H})$. This suggests that it may be possible for the $\rm PH_{alt}$ model to reproduce the observed emission lines if the ionization rate is increased from our fiducial value of $1.3 \times 10^{-17}\,\rm s^{-1}$. Fig.\,\ref{fig:linesimulations_noabstraction} shows as an example the simulated lines for $\zeta_p({\rm H}) = 5.0 \times 10^{-17}\,\rm s^{-1}$; the core model is otherwise identical to that used in Fig.\,\ref{fig:linesimulations}. The best-fit time in this case is $t = 2.50 \times 10^5 \, \rm yr$. The overall fit is better than the one obtained with the PH model, and in particular the $\rm NH_2D$ o/p ratio is correctly reproduced, but the better agreement comes with a cost. The steep decline of the ammonia abundances at relatively short timescales is incompatible with observations toward the starless core L1544 which show no ammonia depletion even though the core is in an evolved dynamical state \citep{Crapsi07,Caselli17,Sipila19}; it would be rather serendipitous if the observations caught H-MM1 during the short-lived ammonia abundance peak, assuming that the two cores have evolved similarly. This problem may be solved once it is understood how ammonia can maintain a high gas-phase abundance even at high gas densities (see \citealt{Sipila19} for more discussion on this issue). Fig.\,\ref{fig:linesimulations_noabstraction} also shows the simulated lines in the $\rm PH_{alt}$ model for our fiducial cosmic-ray ionization rate, and it is clear that the agreement with the observations is not better as compared to the PH model, except for the fact that the $\rm NH_2D$ o/p ratio corresponds to the observed value.

\begin{figure}
\centering
\includegraphics[width=1.0\columnwidth]{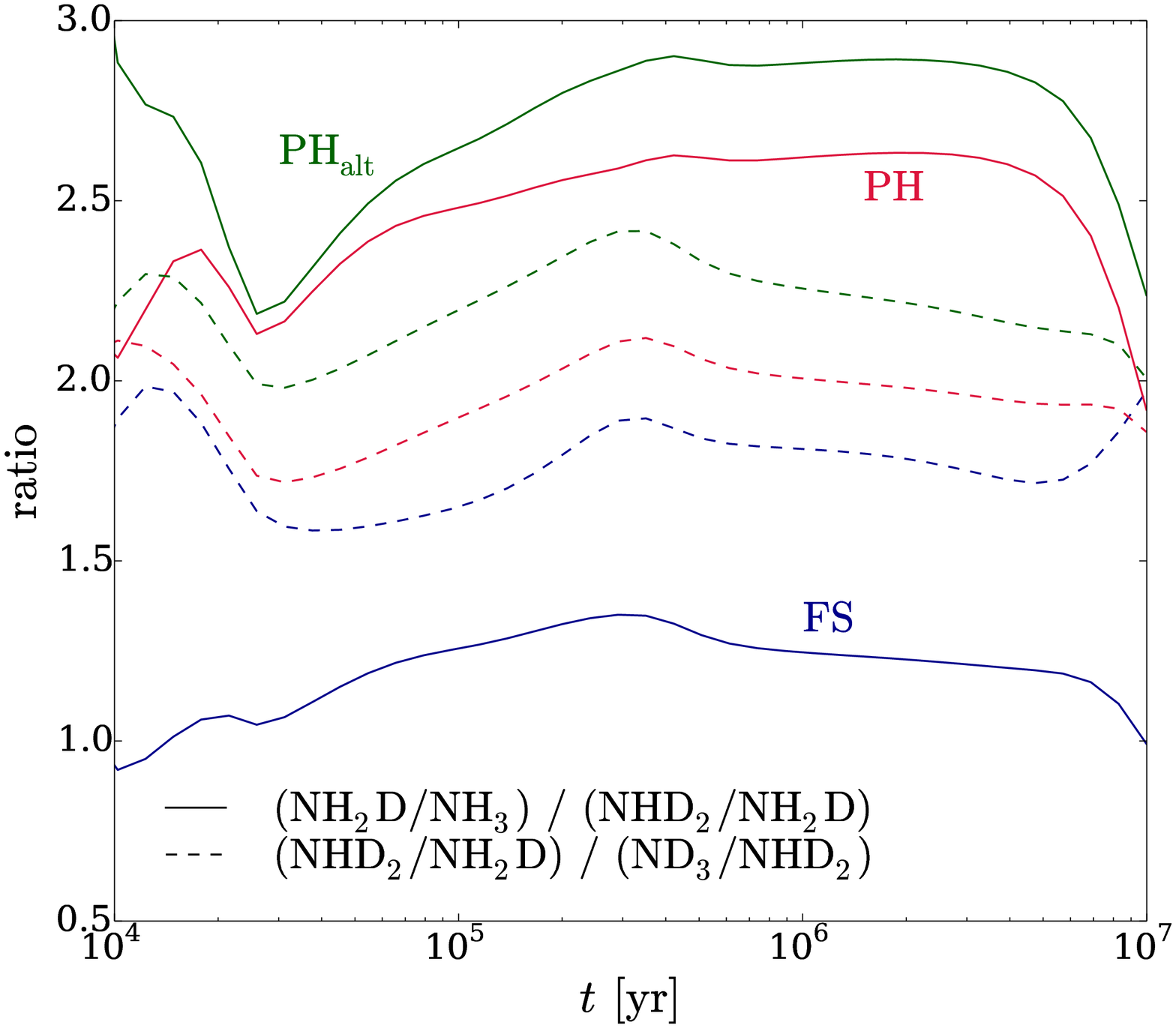}
\caption{Abundance ratios $\rm (NH_2D / NH_3) / (NHD_2 / NH_2D)$ and $\rm (NHD_2 / NH_2D) / (ND_3 / NHD_2)$ as functions of time in the H-MM1 core model. The blue, red, and green curves represent the FS model, the PH model, and the $\rm PH_{alt}$ model, respectively.
}
\label{fig:ammoniaratios}
\end{figure}

As noted in Sect.\,\ref{ss:0dModels}, statistical ammonia deuteration ratios obey $\rm (NH_2D / NH_3) / (NHD_2 / NH_2D) = 3$ and $\rm (NHD_2 / NH_2D) / (ND_3 / NHD_2) = 3$. Model predictions of these ratios depend on the deuterated ammonia formation mechanism \citep{Rodgers02}, and thus they may be useful in discriminating between the FS and PH scenarios. Figure\,\ref{fig:ammoniaratios} shows the time-evolution of these ratios in the H-MM1 core model in the three cases studied above (FS, PH, and  $\rm PH_{alt}$). We recover the same tendency as for point models (Sect.\,\ref{ss:0dModels}), i.e., that the $\rm (NHD_2 / NH_2D) / (ND_3 / NHD_2)$ ratio is higher than the $\rm (NH_2D / NH_3) / (NHD_2 / NH_2D)$ ratio in the FS model, and vice versa for the PH model. Switching from the PH to the $\rm PH_{alt}$ model increases both ratios by a similar factor. These results highlight the strong model-dependent behavior of the $\rm (NH_2D / NH_3) / (NHD_2 / NH_2D)$ ratio in particular. The observed values in H-MM1 are $\rm (NH_2D / NH_3) / (NHD_2 / NH_2D) = 1.77 \pm 0.19$ and $\rm (NHD_2 / NH_2D) / (ND_3 / NHD_2) = 3.67 \pm 0.70$, as calculated from Table~4 in H17. Neither one of these ratios is reproduced by the models. However, we note that the comparison depends on how the column densities are derived; H17 simply assumed a constant abundance throughout the core when determining the best-fit column density. Comparison of line profiles remains the best method of constraining models with the aid of observations, because this allows to take into account the effects of kinematics and optical depth, for example.

We did not consider any corrections to rate coefficients due to differences in mass or zero-point energy between hydrogenated and deuterated species, and so deuterated reactions proceed with the same rate coefficients as their hydrogenated counterparts. We did test the relative rates for the deuterated $\rm NH_3^+ + H_2 \longrightarrow \rm NH_4^+ + H$ reactions tabulated by \citeauthor{Roueff15}\,(\citeyear{Roueff15}; their Table~3), but this change had a negligible influence on our results, which is readily explained by the fact that collisions of ammonia with HD or $\rm D_2$ are relatively rare. Unfortunately, in general little theoretical or laboratory data is available for the deuterated reactions involved in the main ammonia formation pathway; our results show that more work on these reactions is urgently needed to reach a satisfactory agreement between chemical models and observations. The fact that the $\rm PH_{alt}$ model shows such a good agreement with the observations for the $\rm NH_2D$ and $\rm NHD_2$ spin ratios, while simultaneously failing to match the D/H ratios (assuming $\zeta_p({\rm H}) = 1.3 \times 10^{-17}\,\rm s^{-1}$), is especially puzzling. The $\rm NH_3$ o/p ratio has not been measured toward H-MM1 because the low-lying rotational lines (belonging to ortho-$\rm NH_3$) cannot be observed from the ground, but the ratio that the $\rm PH_{alt}$ model predicts ($\sim$1.2) is clearly higher than the value ($\sim$0.7) inferred toward L1544 by \citet{Caselli17}.

It is possible that non-thermal desorption of ammonia from grain surfaces plays a part in setting the abundance ratios in the gas phase. Ammonia is formed on the grain surfaces by sequential association reactions, and in these reactions the products are formed in statistical ratios; for example, $\rm NH + H$ leads to ortho and para $\rm NH_2$ in a 3:1 ratio. Thus, if ammonia was released in significant quantities from the grain surfaces and was mixed with the non-statistical gas-phase distribution, the resultant abundance ratios could be close to statistical as observed by H17. Our chemical model however predicts strong ammonia freeze-out, and standard thermal or non-thermal mechanisms do not lead to appreciable ammonia desorption in starless core conditions as we have recently shown in \citet{Sipila19}. We do not delve into this matter further in this paper, and refer the reader to \citet{Sipila19} for a complete discussion including suggested solutions to the problem.

The results presented in this paper highlight the fact that multiply-deuterated species are very clearly affected by the nature of the proton-transfer process, and that our modeling of the starless core H-MM1 seems to support the PH mechanism as the favored one, but it is difficult to draw definitive conclusions based on the limited data available. It is possible that the prevalence of PH or FS is dependent on the reacting system. We propose that pending further theoretical and laboratory work, a systematic observational programme targeting pairs of (multiply-)deuterated species toward different objects could help to constrain the reaction mechanism, when paired with detailed modeling.

\section{Conclusions}\label{s:conclusions}

We investigated the effect of modifying the reaction mechanism in gas-phase proton-donation reactions on the abundances of (deuterated) ammonia and other key species in the starless core H-MM1, limiting the reactions to proceed through a direct proton hop and disallowing atom-exchange processes. We quantified the difference between the full scrambling and hop models by investigating density-averaged abundance profiles and by carrying out line emission simulations using a radiative transfer code. We also constructed an alternative hop model that includes the treatment of hydrogen atom abstraction reactions as a direct hop process. Our results apply to hydrogenated as well as deuterated species, and the effect of the reaction mechanism on spin-state chemistry is consistently taken into account.

We found that the proton hop mechanism favors the formation of singly-deuterated species over multiply-deuterated ones, because in this case many reaction channels that would lead to multiply-deuterated species in the full-scrambling model are closed off. Switching the reaction mechanism also has implications for spin-state abundance ratios. The proton hop model reproduces somewhat better the $\rm NH_2D / NH_3$, $\rm NHD_2 / NH_2D$, and $\rm ND_3 / NHD_2$ ratios observed toward H-MM1 by \citet{Harju17a} than the full-scrambling model does, but in particular the observed o/p ratio of $\rm NH_2D$ is not reproduced by either model -- both models predict a ratio of $\sim$2 while the observed value is close to the statistical value of~3. We carried out radiative transfer simulations of the observed ammonia emission lines to confirm that these tendencies found in the modeled abundance profiles are present in the lines as well. In the alternative model where also hydrogen abstraction reactions proceed through proton hop, the observed ammonia spin-state abundance ratios are reproduced but the D/H ratios are very strongly overestimated. This problem can be mitigated by considering cosmic-ray ionization ratios higher than our fiducial value of $\zeta_p({\rm H}) = 1.3 \times 10^{-17}\,\rm s^{-1}$, but achieving a good match with the observations would also require fine-tuning the physical and chemical model parameters, which is beyond the scope of the present work.

Chemical species produced directly through proton-donation reactions, such as $\rm N_2D^+$ and $\rm DCO^+$ are unaffected by the nature of the proton-donation reaction mechanism. However, species such as $\rm H_2CO$, $\rm H_2O$, and $\rm CH_3$ whose gas-phase formation pathways depend on proton-donation reactions, are strongly affected by the reaction mechanism.

Our models cannot constrain definitively whether the proton hop process prevails over full scrambling (i.e., allowing for proton exchange) at low temperature in the interstellar medium. The results presented here do however present a preference for the former, at least for ammonia. More theoretical and laboratory work on multiple reacting systems is required to shed more light on this problem. In the meantime, we propose that observations of multiple pairs of (multiply-)deuterated species toward different starless and pre-stellar cores could provide constraints for models of low-temperature deuterium and spin-state chemistry.

\begin{acknowledgements}
We thank Stephan Schlemmer for insightful comments on an early version of the manuscript, and the anonymous referee for a constructive report that improved the paper.
\end{acknowledgements}

\bibliographystyle{aa}
\bibliography{ScramblingVsHop.bib}

\end{document}